# Proposing A Symmetric Key Bit-Level Block Cipher

Sarbajit Manna, Saurabh Dutta

*Abstract*— **A novel bit level block cipher based symmetric key cryptographic technique using G.C.D is proposed in this research paper. Entire plain text file is read one character at a time and according to the binary representation of ASCII value of the characters, entire plain text file is divided into n number of 16 bit blocks. Then an agreed-upon symmetric key file is formed by dividing each 16 bit block into two 8 bit sub blocks and by using Greatest Common Divisor (G.C.D) operation among them. The key size is 40 times the number of 16 bit blocks present in the plain text file as each block produces a key of size 40 bits. The character corresponding to the G.C.D value for each block is stored in the cipher text file which is sent by the sender to the receiver along with the symmetric key file. From the symmetric key file and the cipher text file, the receiver by applying reverse procedure, gets back the original plain text file. This technique has several merits, some of them are formation of symmetric key file dynamically, achievement of 50% compression rate in the cipher text file, better security in terms of brute force attack and applicability of the technique for a large number of files of different size and type.**

## I. INTRODUCTION

IN symmetric key cryptography, sender and receiver of a message agrees upon a shared key. The sender uses the shared key to encrypt the message which is sent to the receiver over the network along with the key. The receiver decrypts the encrypted message using the key to get back the original message. The problem in symmetric key cryptography is that if n pairs of parties are involved in the communication, then n pairs of agreed-upon symmetric key are needed [8]-[10].

The main objective of this technique is to safeguard the plain text from brute force attack and to have a compressed version of the cipher text, so a symmetric key file instead of a symmetric key is formed and the cipher text file is formed which is half in size of plain text file. The algorithm consists of following four major components.

Sarbajit Manna is associated as an Assistant Professor with Ramakrishna Mission Vidyamandira, Belur Math, Howrah- 711202, West Bengal, India (Phone: +91-9474339952; e-mail: sarbajitonline@gmail.com).
Dr. Saurabh Dutta is associated as a Professor with Dr. B. C. Roy Engineering College, Durgapur- 713206, West Bengal, India (Phone: +91-9433411450; e-mail: saurabh.dutta@bcrec.ac.in).

A. *Subdivision of Plain Text (PT) File into Blocks*
   Plain Text File ⟶ Block Divided Plain Text File

B. *Symmetric Key File Generation*
   Block Divided Plain Text File ⟶ Symmetric Key File

C. *Encryption Mechanism*
   G.C.D
   Block Divided Plain Text File  Cipher Text File

D. *Decryption Mechanism*
   G.C.D
   Cipher Text File  Plain Text File

Encryption as well as decryption technique is described elaborately in section II. An example implementation of the proposed technique is shown in tabular form in section III. The result and analysis are shown in section IV which displays required analysis parameters and charts. Conclusion along with further scope of work is contained in section V.

## II. PROPOSED ALGORITHM

Greatest Common Divisor (G.C.D) of two or more integers is the largest positive integer that divides the numbers without a remainder. The G.C.D of 14 and 12 is 2 as 2 is the largest positive integer that divides the numbers without a remainder [1],[7].

Encryption and decryption mechanisms are described in section A and section B respectively.

A. *Encryption Mechanism*

Following four steps describe the encryption mechanism in sequential manner.

Binary stream of bits formation of Plain Text (PT) file

Subdivision of plain text file into blocks
↓
Symmetric key file generation from the plain text file
↓
Cipher Text (CT) file generation

Section A.1 to section A.4 describes the steps.

*A.1. Binary stream of bits formation of Plain Text (PT) file*

Each character present in the plain text file is read and according to the ASCII value of each character, plain text file is converted into a file which contains binary stream of bits for the corresponding characters.

Let, '2' and 'd' be two characters read from plain text file. The binary representation of '2' and 'd' will be 00110010 and 01100100 as '2' and 'd' are 50 and 100 in ASCII.

*A.2. Subdivision of plain text file into blocks*

The binary stream of bits of the plain text file is divided into n number of 16 bit size blocks.

*A.3. Symmetric Key file generation from the plain text file*

This technique uses a 40 bit symmetric key having 5 blocks of length 8 bits each for each 16 bit block. Each 16 bit block is again subdivided into two 8 bit sub blocks. The $1^{st}$ 8 bit block of each 40 bit symmetric key represents sum of positional weights of zero valued bit positions of the $1^{st}$ 8 bit sub block. The $2^{nd}$ 8 bit block of each 40 bit symmetric key represents sum of positional weights of zero valued bit positions of the $2^{nd}$ 8 bit sub block. The $3^{rd}$ 8 bit block of each 40 bit symmetric key represents the Remaining Product (RP) of $1^{st}$ 8 bit sub block. The $4^{th}$ and the $5^{th}$ 8 bit block of each 40 bit symmetric key contain the quotient and the remainder respectively when RP of $2^{nd}$ 8 bit sub block is divided by the RP of $1^{st}$ 8 bit sub block. All entries in the key are stored in binary form. The term 'RP' is described as follows. If any two numbers are divided by their G.C.D, then the quotients in each case are the RPs for the numbers.

So, the plain text block corresponding to the characters '2' and 'd' is 0011001001100100, which is subdivided into two 8 bit sub blocks 00110010 and 01100100 (ASCII value 50 and 100 respectively). The $1^{st}$ and the $2^{nd}$ block of the symmetric key for this 16 bit block will contain 128+64+8+4+1 = 205 = 11001101 and 128+16+8+2+1 = 155 = 10011011 respectively. As the G.C.D between 50 and 100 is 50, so RP for $1^{st}$ and $2^{nd}$ 8 bit block will be 1 and 2 respectively. When 2 is divided by 1 produces quotient and remainder as 2 and 0 respectively. So 00000001, 00000010 and 00000000 are stored in $3^{rd}$, $4^{th}$ and $5^{th}$ block of the symmetric key for this 16 bit block.

*A.4. Cipher Text (CT) generation*

The cipher text character corresponding to each 16 bit block is the G.C.D between two 8 bit sub blocks. Cipher text file is formed by calculating G.C.D for each 16 bit sub block and storing each cipher text character. If there are odd number of characters present in the plain text file, then a block of size 8 bits will remain as an extra block. So to calculate G.C.D for this block, the $2^{nd}$ number is also taken as the decimal equivalent of the extra 8 bit block.

So, cipher text character corresponding to the plain text characters '2' and 'd' will be '2' as G.C.D between 50 and 100 (ASCII value of '2' and 'd' respectively) is 50.

*B. Decryption Mechanism*

Following three steps represents the decryption mechanism.
Conversion of Cipher Text (CT) file in binary form
↓
Subdivision of cipher text file into blocks
↓
Plain text file generation
Steps are described in section B.1 to section B.3.

*B.1. Conversion of Cipher Text (CT) file in binary form*

Each character present in the cipher text file is read and according to the ASCII value of each character, cipher text file is converted into a file which contains binary stream of bits for the corresponding characters.

Let, '2' be a character read from the cipher text file. The binary representation of '2' will be 00110010 as '2' in ASCII is 50.

*B.2. Subdivision of cipher text file into blocks*

The binary stream of bits of the cipher text file is divided into n number of 8 bit size blocks.

*B.3. Plain text file generation*

The decimal equivalent of each cipher text character corresponding to each 8 bit block is calculated which is the G.C.D. $3^{rd}$ block of the symmetric key contains the RP of the $1^{st}$ 8 bit sub block. When it is multiplied with the G.C.D, gives the $1^{st}$ number. $4^{th}$ and $5^{th}$ block of the symmetric key contains the quotient and the remainder respectively when RP of $2^{nd}$ 8 bit sub block is divided by the RP of $1^{st}$ 8 bit sub block. When RP of the $1^{st}$ 8 bit sub block is multiplied with the quotient and then added with the remainder, gives the RP of the $2^{nd}$ 8 bit sub block. When it is multiplied with the G.C.D, gives the $2^{nd}$ number. This process is repeated for each character of the cipher text file to generate the plain text file.

So, decimal equivalent of cipher text character '2' is 50. The RP of the $1^{st}$ 8 bit sub block is read from the $3^{rd}$ block of the symmetric key which is 1 in decimal. The quotient and the remainder are read from the $4^{th}$ and the $5^{th}$ block of the symmetric key which are 2 and 0 in decimal respectively. 1*2+0 = 2, gives the RP of the $2^{nd}$ 8 bit sub block. 1 (RP of the $1^{st}$ 8 bit sub block) is multiplied with 50 (G.C.D) to get the $1^{st}$ 8 bit sub block as 50. 2 (RP of the $2^{nd}$ 8 bit sub block) is multiplied with 50 (G.C.D) to get the $2^{nd}$ 8 bit sub block as 100.

III. IMPLEMENTATION

This section contains two tables. The encryption process of the proposed algorithm is described in Table I and the decryption process of the proposed algorithm is described in Table II. Let, ''in'' be the content of a plain text file of size 1 KB (test.txt) which is taken for encryption. The encrypted file ct_test.txt is formed from the plain text file test.txt. From ct_test.txt, the decrypted file pt_test.txt is formed by decryption. The contents of the plain text file test.txt and the decrypted file pt_test.txt are compared to check whether they are same and it is seen that the contents are indeed same.

TABLE I
ENCRYPTION PROCESS

| Plain Text (PT) | i (ASCII : 105) | n (ASCII : 110) |
|---|---|---|
| Binary equivalent of ASCII value | 01101001 | 01101110 |
| G.C.D of two 8 bit block | 5 (00000101) | |
| Sum of positional weights of zero valued bit positions (First 2 blocks of key) | 150 (10010110) | 145 (10010001) |
| Remaining Product (RP) | 21 (00010101) (3rd block of key) | 22 (00010110) |
| Quotient (4th block of key) | 1 (00000001) | |
| Remainder (5th block of key) | 1 (00000001) | |
| Cipher Text (CT) | ENQ (Control Character) (ASCII : 5) | |

TABLE II
DECRYPTION PROCESS

| Cipher Text (CT) | ENQ (Control Character) (ASCII : 5) | |
|---|---|---|
| G.C.D | 5 | |
| Remaining Product (RP) (3rd block of key) | 21 (00010101) | |
| Quotient (4th block of key) | 1 (00000001) | |
| Remainder (5th block of key) | 1 (00000001) | |
| ASCII value of Plain Text (PT) characters | 21 * 5 = 105 (RP * G.C.D) | (21 * 1 + 1) * 5 = 110 (RP * Quotient + Remainder) * G.C.D |
| Plain Text (PT) | i (ASCII : 105) | n (ASCII : 110) |

## IV. RESULT AND ANALYSIS

.EXE, .DLL, .COM, .SYS, and .TXT are five different types of files which are taken for result and analysis purpose. Ten files of different name, size and content from each type are taken. There are one table for each file type which contains information on source file size, target file size, encryption time, chi square value with the degree of freedom and avalanche percentage. The achieved compression rate in the cipher text file in each case is 50%. The proposed technique has been implemented using C language on a computer with Intel Pentium IV 2.40 GHz processor having 512 MB RAM.

To test the non-homogeneity between source and encrypted file, Pearson's chi-squared test has been performed. It means whether the observations onto encrypted files are in good agreement with a hypothetical distribution. In this case, the chi square distribution is being performed with (256-1)=255 degrees of freedom, 256 being the total number of classes of possible characters in the source as well as in the encrypted files. If the observed value of the statistic exceeds the tabulated value at a given level, the null hypothesis is rejected [2]-[3].

The "Pearson's chi-squared" or the "Goodness-of-fit chi-square" is defined by the following equation:

$$X^2 = \Sigma \{(f_0 - f_e)^2 / f_e\} \quad (1)$$

Here $f_e$ and $f_0$ respectively stand for the frequency of a character in the source file and that of the same character in the corresponding encrypted file. The chi-square values have been calculated on the basis of this formula for each pairs of source and encrypted files [2]-[3].

Avalanche effect is a very important property of any cryptographic algorithm. It is evident if, when an input is changed slightly, the output changes drastically. It states that, if a bit in the plain text is flipped or changed, then almost half of the cipher text bits are changed in the cipher text. The small change can occur either in the plaintext or in the key or both so that it can cause a drastic change in the cipher text. The 5th bit of each plaintext sub block is flipped in the proposed technique. The key for each plaintext block changes dynamically with the flipping of bits in the plaintext block. The avalanche percentage for each file is shown in following tables [4]-[6].

The result for .EXE, .DLL, .COM, .SYS, .TXT files is described in section A.1, section A.2, section A.3, section A.4 and section A.5 respectively.

### A.1. Result for .EXE files

Different analysis parameters for ten .EXE files are presented in Table III. The file sizes vary from 5632 bytes to 21811 bytes. The size of encrypted file, decrypted file and the source file is same. The encryption time varies from 63.22 seconds to 267.21 seconds. The decryption time varies from 61.74 seconds to 215.09 seconds. The chi square value lies in the range 1185 to 7962 with the degree of freedom ranging from 178 to 255. The achieved avalanche is in the range from 7.72% to 15.82%. The compression rate for all .EXE files is 50%.

TABLE III
RESULT FOR .EXE FILES

| Source/ Target File Size (Byte) | Encryption Time (S) | Decryption Time (S) | Chi Square Value | Degree of Freedom | Avalanche Achieved (%) |
|---|---|---|---|---|---|
| 12288 | 158.62 | 137.15 | 2429 | 178 | 13.30 |
| 11776 | 136.55 | 128.14 | 5832 | 249 | 11.37 |
| 11264 | 136.59 | 127.75 | 5997 | 245 | 11.07 |
| 8192 | 100.13 | 92.87 | 1185 | 248 | 11.96 |
| 5632 | 63.22 | 61.74 | 2166 | 217 | 14.34 |
| 7680 | 91.83 | 86.50 | 5770 | 233 | 11.76 |
| 18432 | 217.61 | 215.09 | 7805 | 255 | 10.66 |
| 15360 | 187.95 | 182.57 | 3480 | 253 | 15.82 |
| 21811 | 267.21 | 112.99 | 7962 | 255 | 7.72 |
| 12288 | 162.14 | 139.57 | 2191 | 199 | 13.21 |

The column chart in Fig. 1 graphically compares the encryption time with the source file size for .EXE files. Each gray horizontal bar of series 1 represents the source file size in KB and black horizontal bar of series 2 represents the encryption time in second. It is seen that encryption time varies directly proportionally to the source file size.

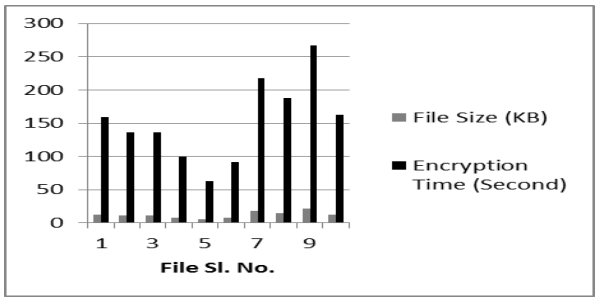

Fig. 1 A comparison between source file size with encryption time and decryption time for .EXE files

*A.2. Result for .DLL files*

Different analysis parameters for ten .DLL files are presented in Table IV. The file sizes vary from 10752 bytes to 34816 bytes. The size of encrypted file, decrypted file and the source file is same. The encryption time varies from 90.74 seconds to 404.91 seconds. The decryption time varies from 58.49 seconds to 269.85 seconds. The chi square value lies in the range 226 to 10896 with the degree of freedom ranging from 148 to 255. The achieved avalanche is in the range from 8.60% to 23.80%. The compression rate for all .DLL files is 50%.

TABLE IV
RESULT FOR .DLL FILES

| Source/ Target File Size (Byte) | Encryption Time (S) | Decryption Time (S) | Chi Square Value | Degree of Freedom | Avalanche Achieved (%) |
|---|---|---|---|---|---|
| 17408 | 187.46 | 127.48 | 6182 | 255 | 10.39 |
| 17408 | 90.74 | 58.49 | 226 | 148 | 23.80 |
| 13312 | 133.85 | 95.9 | 2469 | 246 | 17.45 |
| 30208 | 344.87 | 239.53 | 8442 | 255 | 10.84 |
| 34816 | 404.91 | 269.85 | 8465 | 255 | 11.35 |
| 11264 | 125.23 | 80.63 | 6080 | 252 | 11.24 |
| 16896 | 185.15 | 131.06 | 10896 | 245 | 8.60 |
| 10752 | 118.69 | 81.95 | 3864 | 250 | 12.96 |
| 15872 | 172.52 | 116.77 | 6375 | 254 | 11.02 |
| 13312 | 142.15 | 96.45 | 2877 | 252 | 15.50 |

The column chart in Fig. 2 graphically compares the encryption time with the source file size for .DLL files. Each gray horizontal bar of series 1 represents the source file size in KB and black horizontal bar of series 2 represents the encryption time in second. It is seen that encryption time varies directly proportionally to the source file size.

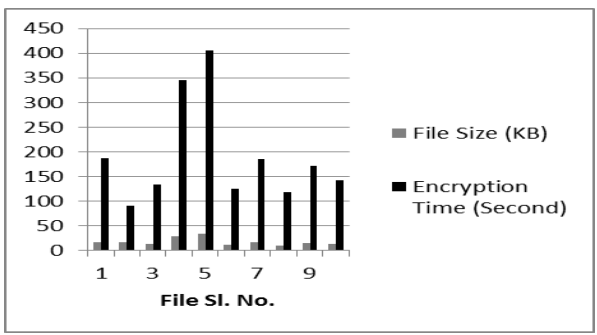

Fig. 2 A comparison between source file size with encryption time and decryption time for .DLL files

*A.3. Result for .COM files*

Different analysis parameters for ten .COM files are presented in Table V. The file sizes vary from 1126 bytes to 69836 bytes. The size of encrypted file, decrypted file and the source file is same. The encryption time varies from 10.32 seconds to 960.35 seconds. The decryption time varies from 6.48 seconds to 395.46 seconds. The chi square value lies in the range 1010 to 11431 with the degree of freedom ranging from 148 to 255. The achieved avalanche is in the range from 6.56% to 12.48%. The compression rate for all .COM files is 50%.

TABLE V
RESULT FOR .COM FILES

| Source/ Target File Size (Byte) | Encryption Time (S) | Decryption Time (S) | Chi Square Value | Degree of Freedom | Avalanche Achieved (%) |
|---|---|---|---|---|---|
| 7680 | 89.59 | 77.5 | 1939 | 238 | 11.92 |
| 9216 | 109.31 | 95.3 | 3593 | 245 | 11.53 |
| 7168 | 82.28 | 74.97 | 6415 | 233 | 12.48 |
| 69836 | 960.35 | 395.46 | 11431 | 255 | 6.93 |
| 29696 | 351.47 | 299.07 | 8595 | 254 | 9.72 |
| 26112 | 313.57 | 311.59 | 8582 | 255 | 9.75 |
| 19660 | 228.65 | 128.85 | 8900 | 254 | 7.71 |
| 7014 | 78.54 | 39.32 | 5794 | 255 | 7.11 |
| 14643 | 159.51 | 82.88 | 7624 | 255 | 7.27 |
| 1126 | 10.32 | 6.48 | 1010 | 148 | 6.56 |

The column chart in Fig. 3 graphically compares the encryption time with the source file size for .COM files. Each gray horizontal bar of series 1 represents the source file size in KB and black horizontal bar of series 2 represents the encryption time in second. It is seen that encryption time varies directly proportionally to the source file size.

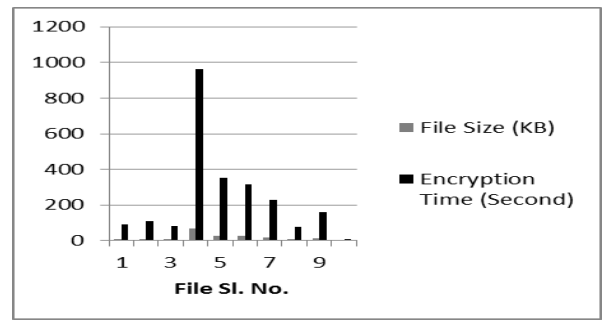

Fig. 3 A comparison between source file size with encryption time and decryption time for .COM files

*A.4. Result for .SYS files*

Different analysis parameters for ten .SYS files are presented in Table VI. The file sizes vary from 4218 bytes to 17612 bytes. The size of encrypted file, decrypted file and the source file is same. The encryption time varies from 43.78 seconds to 188.78 seconds. The decryption time varies from 27.63 seconds to 117.65 seconds. The chi square value lies in the range 2530 to 6838 with the degree of freedom ranging from 208 to 255. The achieved avalanche is in the range from 6.91% to 15.19%. The compression rate for all .SYS files is 50%.

TABLE VI
RESULT FOR .SYS FILES

| Source/Target File Size (Byte) | Encryption Time (S) | Decryption Time (S) | Chi Square Value | Degree of Freedom | Avalanche Achieved (%) |
|---|---|---|---|---|---|
| 10240 | 117.87 | 101.89 | 4170 | 252 | 11.40 |
| 11264 | 135.44 | 90.46 | 4861 | 247 | 11.52 |
| 15872 | 164.45 | 117.65 | 6838 | 255 | 10.25 |
| 10240 | 106.78 | 83.92 | 4208 | 247 | 11.34 |
| 10240 | 111.66 | 80.41 | 3988 | 250 | 11.34 |
| 9021 | 82.11 | 55.04 | 4782 | 251 | 7.88 |
| 4218 | 43.78 | 27.63 | 2530 | 208 | 12.67 |
| 16384 | 188.78 | 115.29 | 4560 | 241 | 11.15 |
| 4761 | 44.10 | 29.33 | 3550 | 241 | 6.91 |
| 17612 | 185.54 | 116.17 | 4421 | 253 | 15.19 |

The column chart in Fig. 4 graphically compares the encryption time with the source file size for .SYS files. Each gray horizontal bar of series 1 represents the source file size in KB and black horizontal bar of series 2 represents the encryption time in second. It is seen that encryption time varies directly proportionally to the source file size.

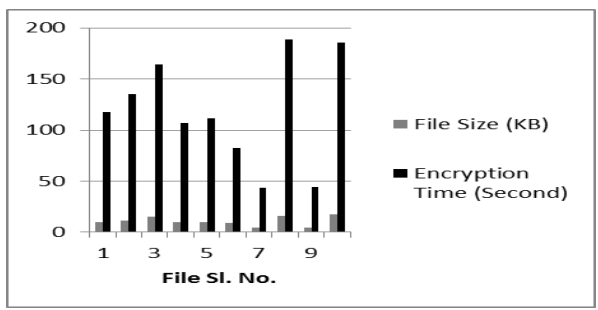

Fig. 4 A comparison between source file size with encryption time and decryption time for .SYS files

*A.5. Result for .TXT files*

Different analysis parameters for ten .TXT files are presented in Table VII. The file sizes vary from 133 bytes to 32563 bytes. The size of encrypted file, decrypted file and the source file is same. The encryption time varies from 0.94 seconds to 205.15 seconds. The decryption time varies from 0.22 seconds to 52.40 seconds. The chi square value lies in the range 1 to 5706 with the degree of freedom ranging from 35 to 105. The achieved avalanche is in the range from 5.60% to 27.88%. The compression rate for all .TXT files is 50%.

TABLE VII
RESULT FOR .TXT FILES

| Source/Target File Size (Bytes) | Encryption Time (S) | Decryption Time (S) | Chi Square Value | Degree of Freedom | Avalanche Achieved (%) |
|---|---|---|---|---|---|
| 1331 | 8.13 | 2.20 | 1205 | 78 | 7.75 |
| 5416 | 35.76 | 9.33 | 1974 | 91 | 7.35 |
| 6451 | 41.91 | 9.78 | 792 | 88 | 10.98 |
| 133 | 0.94 | 0.22 | 125 | 35 | 5.60 |
| 829 | 5.17 | 1.26 | 690 | 65 | 9.41 |
| 22937 | 146.76 | 37.13 | 1707 | 100 | 10.66 |
| 2201 | 13.79 | 3.63 | 1569 | 80 | 7.91 |
| 2498 | 15.38 | 4.23 | 1641 | 79 | 7.98 |
| 32563 | 205.15 | 52.40 | 1 | 61 | 27.88 |
| 19046 | 119.47 | 28.45 | 5706 | 105 | 8.56 |

The column chart in Fig. 5 graphically compares the encryption time with the source file size for .TXT files. Each gray horizontal bar of series 1 represents the source file size in KB and black horizontal bar of series 2 represents the encryption time in second. It is seen that encryption time varies directly proportionally to the source file size.

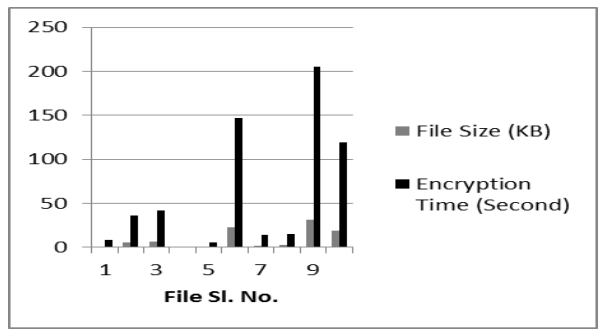

Fig. 5 A comparison between source file size with encryption time and decryption time for .TXT files

## V. CONCLUSION

The following factors determine the merits of the technique. The first among them is to make a dynamic symmetric key file which is directly derived from the plain text blocks. The second factor is the security in terms of brute force attack. To enhance security, a symmetric key file is generated and it is sent to the recipient instead of the symmetric key for each block. Thirdly, to reduce overhead on the network, a 50% compressed version of the cipher text file is generated Half memory space is needed to store the encrypted file. The symmetric key file size is totally dependent on the source file size. The more the size of the source file, the more is the symmetric key file size. The technique can be equally implemented in any popular high level language. It is simple to implement. As the encryption and decryption has been done in bit level, the execution time is dependent on the source file size. The achieved avalanche percentage is between 10 and 16 for most of the files. The avalanche percentage will be much better if more than one bit of the plain text file is flipped. By flipping any one or more bits of the plain text file or by flipping any one or more number of bits of the symmetric key file or even both the plain text file and the symmetric key file, the achieved avalanche percentage can also be calculated. The cipher text file can also be formed not only using G.C.D operation but also any one or more reversible logical operations.